\documentclass[10pt,superscriptaddress,nofootinbib,notitlepage,tightenlines,twocolumn, pra]{revtex4-1}

\usepackage{amsthm}
\usepackage{amsmath}
\usepackage{amssymb}
\usepackage{graphicx}
\usepackage{epstopdf}
\usepackage{url}
\usepackage{float}
\usepackage{bm}
\usepackage{ifthen}
\usepackage[usenames,dvipsnames]{color}
\usepackage{mathrsfs}
\usepackage[colorlinks=true,citecolor=blue,urlcolor=black]{hyperref}
\usepackage{float}

\newcommand{\tn}[1]{\textnormal{#1}}
\newcommand{\be}{\begin{equation}}
\newcommand{\ee}{\end{equation}}

\newcommand{\rv}[1]{{\bf{#1}}}

\newcommand{\leak}{\tn{leak}_{\rm EC}}
\newcommand{\epscorr}{\eps_{\tn{cor}}}
\newcommand{\esec}{\eps_{\tn{sec}}}

\newcommand{\sket}[1]{{\ensuremath{\lvert#1\rangle}}}
\newcommand{\lket}[1]{{\ensuremath{\left\lvert#1\right\rangle}}}
\newcommand{\ket}[1]{\if@display\lket{#1}\else\sket{#1}\fi}

\newcommand{\sbra}[1]{{\ensuremath{\langle#1\rvert}}}
\newcommand{\lbra}[1]{{\ensuremath{\left\langle#1\right\rvert}}}
\newcommand{\bra}[1]{\if@display\lbra{#1}\else\sbra{#1}\fi}

\newcommand{\sbraket}[2]{{\ensuremath{\langle#1\rvert#2\rangle}}}
\newcommand{\lbraket}[2]{{\ensuremath{\left\langle#1\!\left\rvert\vphantom{#1}#2\right.\!\right\rangle}}}
\newcommand{\braket}[2]{\if@display\lbraket{#1}{#2}\else\sbraket{#1}{#2}\fi}

\newcommand{\sketbra}[2]{{\ensuremath{\lvert #1\rangle\!\langle #2\rvert}}}
\newcommand{\lketbra}[2]{{\ensuremath{\left\lvert #1\right\rangle\!\!\left\langle #2\right\rvert}}}
\newcommand{\ketbra}[2]{\if@display\lketbra{#1}{#2}\else\sketbra{#1}{#2}\fi}

\newcommand{\eps}{\varepsilon}

\newcommand{\cX}{\mathcal{X}}
\newcommand{\cK}{\mathcal{K}}

\newcommand{\cZ}{\mathcal{Z}}

\newcommand{\Eve}{\textbf{E}}
\newcommand{\rvZ}{\textbf{Z}}

\newcommand{\sX}{\mathsf{X}}
\newcommand{\sZ}{\mathsf{Z}}

\newcommand{\rs}{{\rm{1}}}
\newcommand{\rd}{{\rm{2}}}
\newcommand{\rdd}{{\rm{3}}}

\newcommand{\rvX}{\textbf{X}}

\newcommand{\HmaxOp}{H_{\max}}
\newcommand{\HminOp}{H_{\min}}

\theoremstyle{plain}

\theoremstyle{definition}

\begin{document}
\title{Concise Security Bounds for Practical Decoy-State Quantum Key Distribution }
\author{Charles Ci Wen \surname{Lim}}
\email{ciwen.lim@unige.ch}
\affiliation{Group of Applied Physics, University of Geneva, Switzerland.}
\author{Marcos Curty}
\affiliation{EI Telecomunicaci\'on, Dept. of Signal Theory and Communications, University of Vigo, Spain}
\author{Nino \surname{Walenta}}
\affiliation{Group of Applied Physics, University of Geneva, Switzerland.}
\author{Feihu Xu}
\affiliation{Center for Quantum Information and Quantum Control,
Dept. of Physics and Dept. of Electrical \& Computer Engineering,
University of Toronto, Canada }
\author{Hugo \surname{Zbinden}}
\affiliation{Group of Applied Physics, University of Geneva, Switzerland.}

\begin{abstract}
Due to its ability to tolerate high channel loss, decoy-state quantum key distribution (QKD) has been one of the main focuses within the QKD community. Notably, several experimental groups have demonstrated that it is secure and feasible under real-world conditions. Crucially, however, the security and feasibility claims made by most of these experiments were obtained under the assumption that the eavesdropper is restricted to particular types of attacks or that the finite-key effects are neglected.~Unfortunately, such assumptions are not possible to guarantee in practice. In this work, we provide concise and tight finite-key security bounds for practical decoy-state QKD that are valid against general attacks.
\end{abstract} 
\maketitle

{\bf{Introduction.}}~In 1984, Bennett and Brassard proposed a quantum key distribution (QKD) scheme in which a cryptographic key can be securely distributed between two remote parties, Alice and Bob, in an untrusted environment~\cite{bb84}.~Since then, this proposal (traditionally referred to as the BB84 protocol) has received considerable attention, and significant progress has been made in both theory and practice~\cite{Gisin2002}. 

In actuality, implementations of the BB84 protocol differ in some important aspects from the original theoretical proposal. This is particularly the case in the choice of the quantum information carrier, where a weak pulsed laser source is used in place of an ideal single-photon source (which is not yet available). However, pulsed laser sources have a critical drawback in that a non-negligible fraction of the emitted laser pulses contain more than one photon, which an adversary, Eve, can exploit via the so-called photon-number-splitting (PNS) attack~\cite{pns}.~In fact, this attack has been shown to be extremely powerful, especially when the loss in the quantum channel connecting Alice and Bob is high.

To tackle the PNS attack in the presence of high channel loss, most BB84 implementations (e.g., see Refs.~\cite{Zhao2006, Rosenberg2007, Peng2007, Yuan2007, Dixon2008, Tanaka2008, Rosenberg2009, Dixon2010, Liu2010, Sasaki2011, Wang2013})~adopt the decoy-state method~\cite{Hwang2003, Lo2005M, Wang2005a}.~The basic idea is conceptually very simple, and more importantly, it requires minimal modification to existing BB84 implementations. Specifically, instead of preparing phase-randomized laser pulses of the same mean photon-number, Alice varies randomly and independently the mean photon-number of each laser pulse she sends to Bob.~Crucially, by using the fact that the variation of the mean photon-number is inaccessible to Eve, it is possible to detect the presence of photon-number-dependent loss in the quantum channel, i.e., by analyzing the data shared between Alice and Bob.~As a result, photon-number-dependent type of attacks are circumvented, and the secret key rates and the tolerance to the channel loss are significantly improved.

The security of decoy-state QKD has been obtained in the asymptotic regime~\cite{Lo2005M, Wang2005a}, i.e., in the limit of infinitely long keys. In the case of finite-length keys, several attempts have been made (e.g., see Refs.~\cite{Hayashi2007b,Cai2009, Song2011,Somma2013}), but, most (if not all) of these results assume that Eve is restricted to particular types of attacks. Very recently, finite-key security bounds against general attacks have been derived by Hayashi and Nakayama~\cite{Hayashi2013}, although the security analysis is rather involved. 

In this work, we provide concise and tight finite-key security bounds for a practical decoy-state QKD protocol that are directly applicable to most current decoy-state QKD implementations.~The security analysis is based on a combination of a recent security proof technique~\cite{Tomamichel2010, Tomamichel2012aM}~and a novel finite-size analysis for the decoy-state method, which allows us to greatly simplify the security analysis.~As a result, we are able to derive tight finite-key security bounds that are valid against general attacks. Moreover, these bounds can be straightforwardly computed with just five concise formulas (see Eqs.~\eqref{eqn1}-\eqref{eqn5}), which experimentalists can readily use for their implementations. In addition, we evaluate the performance of our security bounds by applying them to a realistic fiber-based system model.~The evaluation shows that our security bounds are relatively tight, in the sense that for realistic post-processing block sizes, the achievable secret key rates are comparable to those obtained in the asymptotic regime. In fact, for small post-processing block sizes (of the order of $10^4$ bits), we observe that secret keys can be securely distributed over a fiber length of up to $135$ km. 
 \newline

{\bf{Protocol Description.~}}~We consider an asymmetric coding BB84 protocol~\cite{Lo2005A}, i.e., the bases $\sX$ and $\sZ$ are chosen with probabilities that are biased. Specifically, the bases $\sX$ and $\sZ$ are selected with probabilities $q_{\rm x}$ and $1-q_{\rm x}$, respectively, and the secret key is extracted from the events whereby Alice and Bob both choose the $\sX$ basis. In addition, the protocol is based on the transmission of phase-randomized laser pulses, and uses two-decoy settings. The intensity of each laser pulse is randomly set to one of the three intensities $\mu_1,\mu_2$ and $\mu_3$, and the intensities satisfy $\mu_1> \mu_2+\mu_3$ and $\mu_2 > \mu_3 \geq 0$. Note, however, that our analysis can also be straightforwardly generalized to any number of intensity levels. Next, we provide a detailed description of the protocol. 
\newline

\noindent{\it{1.~Preparation.}}~Alice chooses a bit value uniformly at random and records the value in $y_i$.~Then, she selects a basis choice $a_i \in \{\mathsf{X},\mathsf{Z}\}$ with probabilities $q_{\rm x}$ and $1-q_{\rm x}$, respectively, and an intensity choice $k_i \in \cK:=\{\mu_\rs,\mu_\rd,\mu_\rdd \}$ with probabilities $p_{\mu_\rs}$, $p_{\mu_\rd}$ and $p_{\mu_\rdd}=1-p_{\mu_\rs}-p_{\mu_\rd}$, respectively. Finally, she prepares a (weak) laser pulse based on the chosen values and sends it to Bob via the quantum channel. \newline 

\noindent{\it{2.~Measurement.}}~Bob chooses a basis $b_i\in \{\mathsf{X},\mathsf{Z}\}$ with probabilities $q_{\rm x}$ and $1-q_{\rm x}$, respectively. Then, he performs a measurement in basis $b_i$ and records the outcome in $y_i^\prime$. In practice, the measurement device is usually implemented with two single-photon detectors. In this case, there are four possible outcomes $\{0,1,\emptyset,\perp\}$ where $0$ and $1$ are the bit values, and $\emptyset$ and $\perp$ are the no detection and double detection events, respectively. For the first three outcomes, Bob assigns what he observes to $y^\prime_i$, and for the last outcome $\perp$ he assigns a random bit value to $y^\prime_i$. 
\newline

\noindent{\it{3.~Basis reconciliation.}}~Alice and Bob announce their basis and intensity choices over an authenticated public channel and identify the following sets: $\cX_{k}:=\{i: a_i=b_i=\mathsf{X} \wedge k_i=k  \wedge y_i^\prime \not= \emptyset\}$ and $\cZ_{k}:=\{i: a_i=b_i=\mathsf{Z} \wedge k_i=k  \wedge y_i^\prime \not= \emptyset\}$ for all $k \in \cK$. Then, they check for $|\cX_{k}| \geq n_{\sX,k}$ and $|\cZ_{k}| \geq n_{\sZ,k}$ for all values of $k$. They repeat step 1 to step 3 until
these conditions are satisfied. We denote as $N$ the number of laser pulses sent by Alice until the conditions are fulfilled. 
\newline

\noindent{\it{4.~Generation of raw key and error estimation.}}~First, a raw key pair $(\rv{X}_{\rm A},\rv{X}_{\rm B})$ is generated by choosing a random sample of size $n_{\sX}=\sum_{k\in\cK}n_{\sX,k}$ of $\cX=\cup_{k\in\cK}\cX_{k}$, where $n_\sX$ is the post-processing block size. Note that we use all intensity levels for the key generation, while existing decoy-state QKD protocols typically use only one intensity level. Second, they announce the sets $\cZ_k$ and compute the corresponding number of bit errors, $m_{\sZ,k}$.~Third, they calculate the number of vacuum events ${s}_{\sX,0}$ [Eq.~\eqref{eqn2}] and the number of single-photon events ${s}_{\sX,1}$ [Eq.~\eqref{eqn3}]  in $(\rv{X}_{\rm A},\rv{X}_{\rm B})$. Also, they calculate the number of \emph{phase errors} $c_{\sX,1}$ [Eq.~\eqref{eqn5}] in the single-photon events. Finally, they check that the phase error rate $\phi_\sX$ is less than $ \phi_{\rm{tol}}$ where $\phi_{\rm{tol}}$ is a predetermined phase error rate, $\phi_\sX:=c_{\sX,1}/{s}_{\sX,1} < \phi_{\tn{tol}}$. If this condition is not met, they abort the protocol, otherwise they proceed to step $5$. 
\newline

\noindent{\it{5.~Post-processing.}}~First, Alice and Bob perform an error-correction step that reveals at most $\leak$ bits of information. In this step, we assume that they try to correct for an error rate that is predetermined. 
Next, to ensure that they share a pair of identical keys,~they perform an error-verification step using two-universal hash functions that publishes $\lceil \log_2{1/\eps_\tn{hash}}\rceil$ bits of information~\cite{WC81M}.~Here, $\eps_{\tn{hash}}$ is the probability that a pair of non-identical keys passes the error-verification step. Finally, conditioned on passing this last step, they perform privacy amplification on their keys to extract a secret key pair ($\rv{S}_{\rm A}, \rv{S}_{\rm B}$) where $|\rv{S}_{\rm A}|=|\rv{S}_{\rm B}|=\ell$ bits.
\newline

{\bf{Security bounds.}}~Before we state the security bounds for our protocol, it is instructive to spell out the security criteria that we are using. For some small protocol errors, $\epscorr,\esec>0$, we say that our protocol is $\epscorr+\esec$-secure if it is $\epscorr$-correct and $\esec$-secret. The former is satisfied if $\Pr[\rv{S}_{\rm A}\not=\rv{S}_{\rm B}] \leq \epscorr$, i.e., the secret keys are identical except with a small probability $\epscorr$. The latter is satisfied if $(1-p_{\rm abort})\|\rho_{\rm AE}-U_{\rm A} \otimes \rho_{\rm E}\|_1/2 \leq \esec$ where $\rho_{\rm AE}$ is the classical-quantum state describing the joint state of $\rv{S}_{\rm A}$ and $\rv{E}$,~$U_{\rm A}$ is the uniform mixture of all possible values of $\rv{S}_{\rm A}$,~and $p_{\rm abort}$ is the probability that the protocol aborts.~Importantly, this secrecy criterion guarantees that the protocol is universally composable: the pair of secret keys can be safely used in any cryptographic task, e.g., for encrypting messages, that requires a perfectly secure key~\cite{RennerThesis05}.

In the following, we present only the necessary formulas to compute the security bounds; the full security analysis is deferred to the supplementary material.

~The correctness of the protocol is guaranteed by the error-verification step.~This step ensures that Bob's corrected key is identical to Alice's key with probability at least $1-\eps_\tn{hash}$, which implies that the final secret keys ($\rv{S}_{\rm A}$, $\rv{S}_{\rm B}$) are identical with probability at least $1-\eps_\tn{hash}$. Therefore, the correctness of the protocol is $\epscorr=\eps_\tn{hash}$. 

Conditioned on passing the checks in the error-estimation and error-verification steps, a $\esec$-secret key of length 
\begin{multline}\label{eqn1}
\ell = \bigg\lfloor s_{\sX,0}+s_{\sX,1} -s_{\sX,1}{h}\left( \phi_\sX \right)\\-\leak-6\log_2\frac{21}{\esec}-\log_2\frac{2}{\epscorr}\bigg\rfloor,
\end{multline} can be extracted, where~$h(x):=-x\log_2x-(1-x)\log_2(1-x)$~is the binary entropy function. Recall that~$s_{\sX,0}$, $s_{\sX,1}$ and $\phi_\sX=c_{\sX,1}/s_{\sX,1}$~are the number of vacuum events, the number of single-photon events, and the phase error rate associated with the single-photons events in $\rv{X}_{\rm A}$, respectively. Next, we show how to calculate them in two steps.

First, we extend the decoy-state analysis proposed in Ref.~\cite{Ma2005M} to the case of finite sample sizes. 
Accordingly, the number of vacuum events in $\rv{X}_{\rm A}$ satisfies
\be\label{eqn2}
s_{\sX,0} \geq \tau_{0}\frac{\mu_\rd n_{\sX,\mu_\rdd}^--\mu_\rdd n_{\sX,\mu_\rd}^+}{\mu_\rd-\mu_\rdd},
\ee where $\tau_{n}:=\sum_{k\in\cK}e^{-k}k^np_k/n!$ is the probability that Alice sends a $n$-photon state, and 
\[
n_{\sX,k}^\pm:=\frac{e^{k}}{p_k}\left[n_{\sX,k}\pm\sqrt{ \frac{n_\sX}{2}\log\frac{21}{\esec}}\right],~\forall~k \in \cK.
\]The number of single-photon events in $\rv{X}_{\rm A}$ is
\begin{multline} \label{eqn3}
s_{\sX,1} \geq \frac{\tau_{1}\mu_\rs\left[n_{\sX,\mu_\rd}^--n_{\sX,\mu_\rdd}^+-\frac{\mu_\rd^2-\mu_\rdd^2}{\mu_\rs^2}(n_{\sX,\mu_\rs}^+- \frac{s_{\sX,0}}{\tau_0})\right]}{\mu_\rs(\mu_\rd-\mu_\rdd)-\mu_\rd^2+\mu_\rdd^2}.
\end{multline}
We also calculate the number of vacuum events, $s_{\sZ,0}$, and the number of single-photon events, $s_{\sZ,1}$, for $\cZ=\cup_{k\in\cK}\cZ_k$, i.e., by using Eqs.~\eqref{eqn2} and~\eqref{eqn3} with statistics from the basis $\sZ$. In addition, the number of bit errors $v_{\sZ,1}$ associated with the single-photon events in $\cZ$ is also required. It is given by
\be\label{eqn4}
v_{\sZ,1} \leq \tau_{1}\frac{m_{\sZ,\mu_\rd}^+-m_{\sZ,\mu_\rdd}^-}{\mu_\rd-\mu_\rdd},
\ee
where
\[
m_{\sZ,k}^{\pm}:=\frac{e^{k}}{p_k}\left[m_{\sZ,k}\pm\sqrt{ \frac{m_\sZ}{2}\log\frac{21}{\esec}}\right],~\forall~k \in \cK,
\]
and $m_\sZ=\sum_{k\in\cK} m_{\sZ,k}$.

Second, the formula for the phase error rate of the single-photon events in $\rv{X}_A$ is~\cite{Fung2010M}
\be \label{eqn5}
\phi_\sX:=\frac{c_{\sX,1}}{s_{\sX,1}}  \leq \frac{v_{\sZ,1}}{s_{\sZ,1}} + \gamma\left(\esec, \frac{v_{\sZ,1}}{s_{\sZ,1}},s_{\sZ,1},{s}_{\sX,1} \right), \ee
where
\[
\gamma\left(a,b,c,d \right):= \sqrt{\frac{(c+d)(1-b)b}{cd\log2}\log_2\left( \frac{c+d}{cd(1-b)b} \frac{21^2}{a^2}\right)}.
\]
\newline

{\bf{Evaluation.}}~We consider a fiber-based QKD system model that borrows parameters from recent decoy-state QKD and single-photon detector experiments. In particular, we assume that Alice can set the intensity of each laser pulse to one of the three predetermined intensity levels, $\mu_\rs$, $\mu_\rd$, and $\mu_\rdd=2\times 10^{-4}$~\cite{Frohlich2013}. Bob uses an active measurement setup with two single-photon detectors (InGaAs APDs): they have a detection efficiency of $\eta_{\tn{Bob}}= 10\%$, a dark count probability of $p_{\tn{dc}}= 6\times 10^{-7}$ and an after-pulse probability of $p_{\tn{ap}}=4 \times 10^{-2}$~\cite{Walenta2012}. The measurement has four possible outcomes $\{0,1,\emptyset,\bot \}$ which correspond to bit values 0, 1, no detection and double detection.

\begin{figure}[t!] 
  \centering  
  \includegraphics[width=86mm]{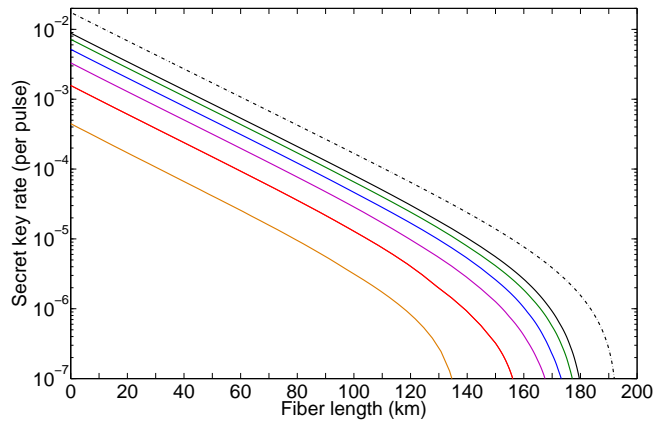}
  \caption{\textbf{Secret key rate vs fiber length (dedicated fiber). }\label{fig1} Numerically optimized secret key rates (in logarithmic scale) are obtained for a fixed post-processing block size $n_{\sX}=10^s$ with $s=4,5,\ldots, 9$ (from left to right).~The dashed curve corresponds to the asymptotic secret key rate, i.e., in the limit of infinitely large keys; however, here we still assume that the number of intensity levels is three. The number of laser pulses sent by Alice can be approximated with the secret key rate and the block size, i.e., $N \leq n_\sX/R$.  }
\end{figure}

The system model is applied to two types of channel architectures, namely one that uses a dedicated optical fiber for the quantum channel and one that uses dense wavelength division multiplexing (DWDM) to put the quantum channel together with the classical channels into one optical fiber (e.g., see Refs.~\cite{Peters2009, Eraerds2010,Patel2012}).~In both cases, we assume that the fibers have an attenuation coefficient of $0.2$ dB/km. That is, their transmittance is~$\eta_{\tn{ch}}=10^{-0.2L/10}$, where $L$ (km) is the fiber length. 

The considered channel architectures, however, do not have the same channel error model. For the dedicated fiber, the probability of having a bit error for intensity $k$ is $e_k=p_{\tn{dc}}+e_{\rm mis}[1-\exp(-\eta_{\tn{ch}}k)]+p_{\tn {ap}}D_k/2$, where $e_{\rm mis}$ is the error rate due to optical errors. Here, the expected detection rate (excluding after-pulse contributions) is $D_{k}=1-(1-2p_{\tn {dc}})\exp(-\eta_{\tn{sys}}k)$, where $\eta_{\tn{sys}}=\eta_\tn{ch}\eta_{\tn{Bob}}$. The expected detection rate (including after-pulse contributions) is thus $R_{k}=D_k(1+p_{\tn{pa}})$. The channel error model for the DWDM architecture is more involved due to additional noise contributions from Raman scattering and cross-talks between channels. We refer to Ref.~\cite{Eraerds2010} for details about it. 

The parameter $\leak$ is set to a simple function $f_{\rm EC}h(e_{\tn{obs}})$ where $f_{\rm EC}$ is the error-correction efficiency and $e_{\tn{obs}}$ is the average of the observed error rates in basis $\sX$ (we note that very recently, a more accurate theoretical model of $\leak$ has been derived in Ref.~\cite{Tomamichel2013}).~In practice, however, $\leak$ should be set to the size of the information exchanged during the error-correction step. Regarding the secrecy, we set $\esec$ to be proportional to the secret key length, that is, $\esec=\kappa\ell$ where $\kappa$ is a security constant; this security constant can be seen as the secrecy leakage per generated bit. 

\begin{figure}[t!] 
  \centering  
  \includegraphics[width=86mm]{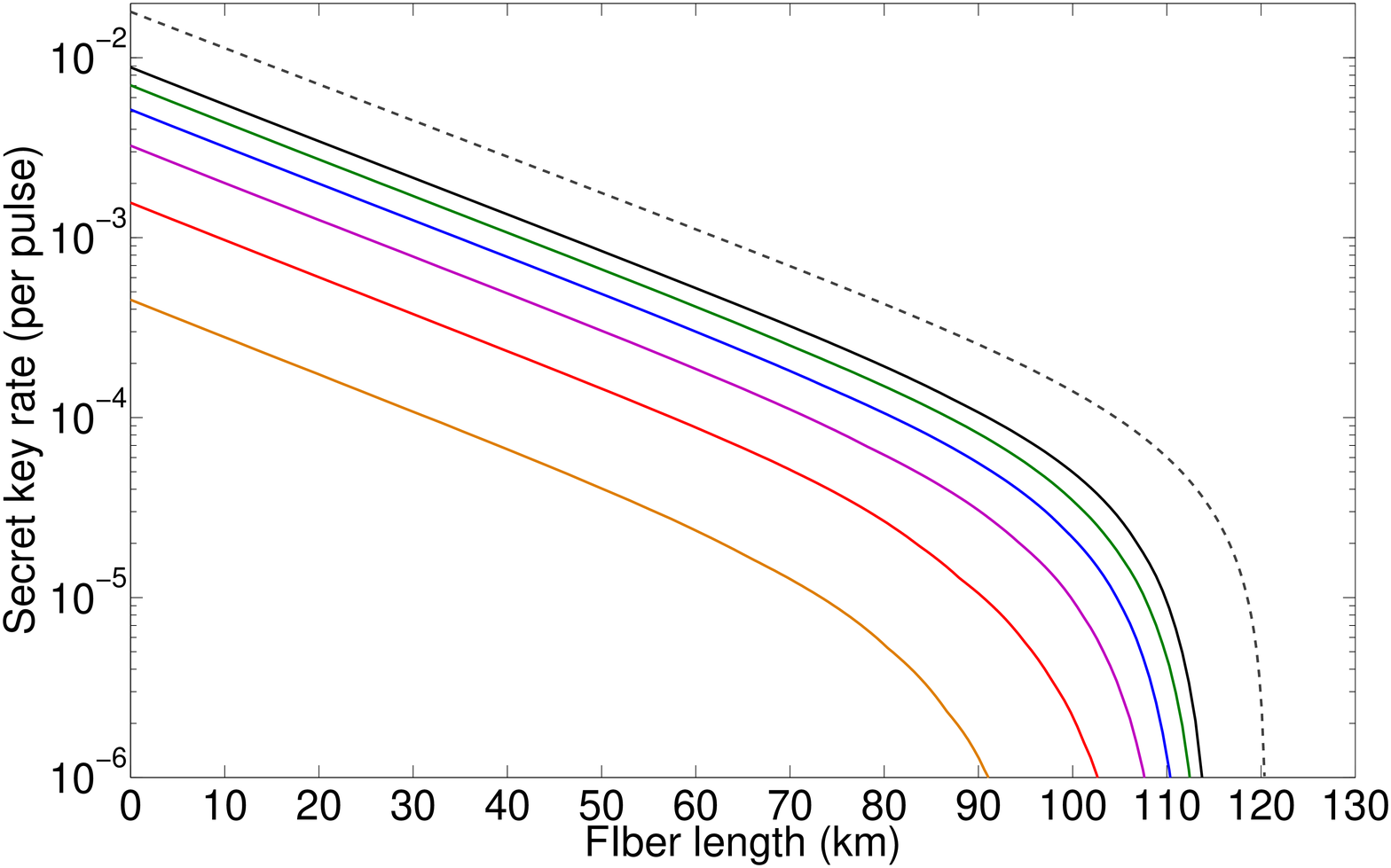}\caption{\textbf{Secret key rate vs fiber length (DWDM).~}\label{fig2} We consider a (4+1) DWDM channel architecture~\cite{Eraerds2010} that puts four classical channels and one quantum channel into an optical fiber. In the simulation, we take that each classical channel has a power of $-34$ dBm at the receiver~\cite{Opticin}. Numerically optimized secret key rates (in logarithmic scale) are obtained for a fixed post-processing block size $n_{\sX}=10^s$ with $s=4,5,\ldots, 9$ (from left to right).~The dashed curve corresponds to the asymptotic secret key rate, i.e., in the limit of infinitely long keys. }
\end{figure} 

For the evaluation, we numerically optimize the secret key rate $R:=\ell/N$ over the free parameters $\{q_{\rm x}, p_{\mu_1}, p_{\mu_2}, \mu_\rs,\mu_\rd\}$~given that the set~$\{\kappa,\epscorr, e_{\tn{mis}}, f_{\rm EC},L,n_\sX\}$ is fixed.~Specifically, we fix $\kappa=10^{-15}$, $\epscorr=10^{-15}$, $e_{\tn{mis}}=5\times 10^{-3}$ and $f_{\rm EC}=1.16$, and generate curves (see Fig.~\ref{fig1}) for a range of realistic post-processing block sizes, i.e., 
$n_{\sX}=10^s$ with $s=4,5,\ldots, 9$.
From Fig.~\ref{fig1}, we see that the security performances corresponding to block sizes $10^7, 10^8$ and $10^9$ have only slight differences. For example, at a fiber length of 100 km, the secret key rate obtained with $n_\sX=10^9$ is about $1.75$ times of the one based on $n_\sX=10^7$. This suggests that it may not be necessary to go to large block sizes (where computational resources are high) to gain significant improvements. On the other hand, for block sizes $10^4, 10^5$ and $10^6$,~there is a distinct advantage in terms of the secret key rate and fiber length for larger block sizes. This is expected since smaller block sizes correspond to larger statistical fluctuations in the estimation process.~Interestingly, we see that even if we use a block size of $10^4$, cryptographic keys can still be distributed over a fiber length of 135 km. The same trend is observed for the DWMD channel architecture (see Fig.~\ref{fig2}). 
\newline

{\bf Concluding remarks.~}Although our security bounds are rather general and can be applied to a wide class of implementations, some conditions on the implementation are still required. In particular, we require that the probability of having a detection in Bob's measurement device is independent of his basis choice.  This condition is normally met when the detectors are operating according to specification. However, if the detectors are not implemented correctly, then there may be serious security consequences, e.g., see Ref~\cite{Lydersen2010}; see also Ref.~\cite{Yuan2011} for the corresponding counter-measures. Alternatively, one can adopt the recently proposed measurement-device-independent QKD (mdiQKD)~\cite{Lo2012} to remove the aforementioned condition; in other words, to remove all detector side-channels. We note, however, that the implementation of mdiQKD is more complex than the one of decoy-state QKD, and the achievable finite-key secret key rates are typically lower~\cite{Curty2013a}.

In summary, we have provided tight finite-key security bounds for a practical decoy-state QKD protocol that can be applied to existing QKD implementations. More importantly, these bounds are secure against general attacks, and can be easily computed by referring to just five concise formulas, i.e., Eqs.~\eqref{eqn1}-\eqref{eqn5}.~On the application side, we also see that secret keys can be securely distributed over large distances with rather small post-processing block sizes.~Accordingly, this allows existing QKD implementations to speed up their key-distillation processes.
\newline

{\bf Acknowledgements.~}We thank Marco Tomamichel and Nicolas Gisin for helpful discussions. We acknowledge support from the Swiss NCCR-QSIT, the NanoTera
QCRYPT, the FP7 Marie-Curie IAAP QCERT project, the European Regional Development Fund (ERDF),
the Galician Regional Government (projects CN2012/279 and CN 2012/260, “Consolidation of Research Units: AtlantTIC”), NSERC, the CRC program, and the Paul Biringer Graduate
Scholarship.

\bibliographystyle{unsrt}

\setcounter{equation}{0}

\section*{Supplementary Material}
Here, we provide the details for the security bounds presented in the main text. The security analysis is a combination of a proof technique based on entropic uncertainty relations~\cite{Tomamichel2011S} and a novel finite-size analysis for the two-decoy-state method. In the following, we first present the details for the decoy-state analysis. 
\newline

\subsection{Decoy-state analysis for three intensity levels}
Recall that our two-decoy-state method consists in Alice setting the intensity of each laser pulse to one of the three intensity levels, $\mu_\rs, \mu_\rd$ and $\mu_\rdd$, where $\mu_\rs > \mu_\rd+\mu_\rdd$ and $\mu_\rd>\mu_\rdd\geq 0$. 
Crucially, from the perspective of the eavesdropper, the final prepared state (i.e., with the encoded bit value) appears the same to her regardless of the choice of intensity level (or equivalently, the average photon-number). Therefore, one can imagine an equivalent \emph{counter-factual} protocol: one in which Alice has the ability to send $n$-photon states, and she only decides on the choice of the average photon-number after Bob has a detection. In the following, we provide the analysis for the $\sX$ basis; the same analysis applies to the $\sZ$ basis.  

Consider the case whereby Alice encodes the states in the $\sX$ basis and let $s_{\sX,n}$~be the number of detections observed by Bob given that Alice sent $n$-photon states. Note that $\sum_{n=0}^\infty s_{\sX,n}=n_{\sX}$ is the total number of detections given that Alice sent states prepared in the $\sX$ basis. In the asymptotic limit, we expect $n_{\sX,k}$ events from $n_\sX$ events to be assigned to the intensity $k$, that is,
\[
n_{\sX,k} \rightarrow n^*_{\sX,k}= \sum_{n=0}^\infty p_{k|n}s_{\sX,n},~\forall k   \in \cK=\{\mu_\rs,\mu_\rd,\mu_\rdd\},
\]
where $p_{k|n}$ is the conditional probability of choosing the intensity $k$ given that Alice prepared a $n$-photon state. For finite sample sizes, using Hoeffding's inequality for independent events~\cite{hoeffding63}, we have that $n_{\sX,k}$ satisfies
\be \label{A:eqn:1}
\left|n^*_{\sX,k}-n_{\sX,k} \right| \leq \delta(n_\sX,\eps_1),
\ee
with probability at least $1-2\eps_1$, where $\delta(n_\sX,\eps_1):=\sqrt{n_\sX/2\log(1/\eps_1)}$. Note that the deviation term $\delta(n_\sX,\eps_1)$ is the same for all values of $k$. Basically, Eq.~\eqref{A:eqn:1} allows us to establish a relation between the asymptotic values and the observed statistics (i.e., $n_{\sX,\mu_\rs}, n_{\sX,\mu_\rd}$ and $n_{\sX,\mu_\rdd}$). Moreover, the same relation can also be made for the expected number of errors and the observed number of errors. Let $v_{\sX,n}$ be the number of errors associated with $s_{\sX,n}$, then in the asymptotic limit, we expect $m_{\sX,k}$ errors from $m_\sX$ errors to be assigned to the intensity $k$, i.e.,
\[
m_{\sX,k} \rightarrow m^*_{\sX,k}=\sum_{n=0}^\infty p_{k|n}v_{\sX,n},~\forall k   \in \cK=\{\mu_\rs,\mu_\rd,\mu_\rdd\}.
\] 
Using Hoeffding's inequality~\cite{hoeffding63}, we thus have for all values of $k$,
\be\label{A:eqn:2}
\left|m^*_{\sX,k}-m_{\sX,k} \right| \leq \delta(m_\sX,\eps_2),
\ee which holds with probability at least $1-2\eps_2$.

For the moment, we keep these relations aside; they will be needed later when we apply the decoy-state analysis (to be detailed below) to the observed statistics.

\subsubsection{Lower-bound on the number of vacuum events}
An analytical lower-bound on $s_{\sX,0}$ can be established by exploiting the \emph{structure} of the conditional probabilities $p_{k|n}$. First of all, we note that with Bayes' rule, for all $k$, we have
\be
p_{k|n}=\frac{p_{k}}{\tau_n}p_{n|k}=\frac{p_{k}}{\tau_n}\frac{e^{-k}k^n}{n!},
\ee
where $\tau_n:=\sum_{k\in\cK}p_{k}e^{-k}k^n/n!$ is the probability that Alice prepares a $n$-photon state. Using this and following an approach proposed by~\cite{Ma2005}, we have that
\begin{multline*}
\frac{\mu_\rd e^{\mu_\rdd} n^*_{\sX,\mu_\rdd}}{p_{\mu_\rdd}}-\frac{\mu_\rdd e^{\mu_\rd}  n^*_{\sX,\mu_\rd}}{p_{\mu_\rd}}\\ =\frac{(\mu_\rd-\mu_\rdd)s_{\sX,0}}{\tau_0} 
- \mu_\rd\mu_\rdd\sum_{n=2}^\infty \frac{(\mu_\rd^{n-1}-\mu_\rdd^{n-1})s_{\sX,n}}{n!\tau_{n}},
\end{multline*}
where the second term on the r.h.s. is non-negative for $\mu_\rd > \mu_\rdd$. Rewriting the above expression for $s_{\sX,0}$ gives
\be \label{A:eqn:4}
s_{\sX,0} \geq \frac{\tau_0}{(\mu_\rd-\mu_\rdd)}\left(\frac{\mu_\rd e^{\mu_\rdd} n^*_{\sX,\mu_\rdd}}{p_{\mu_\rdd}}-\frac{\mu_\rdd e^{\mu_\rd}  n^*_{\sX,\mu_\rd}}{p_{\mu_\rd}}\right).
\ee 
Note that this lower-bound is tight when $\mu_\rdd \rightarrow 0$. 
\subsubsection{Lower-bound on the number of single-photon events }
The lower-bound for the number of single-photon events is slightly more involved, but it can be demonstrated in three concise steps.

First, note that 
\begin{multline*}
\frac{ e^{\mu_\rd}  n^*_{\sX,\mu_\rd}}{p_{\mu_\rd}}-\frac{ e^{\mu_\rdd} n^*_{\sX,\mu_\rdd}}{p_{\mu_\rdd}}\\=\frac{(\mu_\rd-\mu_\rdd)s_{\sX,1}}{\tau_1}+\sum_{n=2}^\infty\frac{(\mu_\rd^n-\mu_\rdd^n)s_{\sX,n}}{n!\tau_n}\\
\leq \frac{(\mu_\rd-\mu_\rdd)s_{\sX,1}}{\tau_1}+\frac{\mu_\rd^2-\mu_\rdd^2}{\mu_\rs^2}   \sum_{n=2}^\infty\frac{\mu_\rs^n s_{\sX,n}}{n!\tau_n},
\end{multline*}
where the inequality is due to 
\begin{multline*}\mu_2^n-\mu_3^n=\frac{(\mu_2^2-\mu_3^2)}{(\mu_2+\mu_3)}\sum_{i=0}^{n-1}\mu_2^{n-i-1}\mu_3^i  \\ \leq (\mu_2^2-\mu_3^2)(\mu_2+\mu_3)^{n-2}  \leq (\mu_2^2-\mu_3^2)\mu_1^{n-2},\end{multline*}
 for $n \geq 2$ and $\mu_2+\mu_3 \leq \mu_1$. Note that we used $\sum_{i=0}^{n-1}\mu_2^{n-i-1}\mu_3^i\leq (\mu_2+\mu_3)^{n-1}$ for $n\geq 2$.

Second, using the fact that the sum of multi-photon events is given by
\[
\sum_{n=2}^\infty\frac{\mu_\rs^n s_{\sX,n}}{n!\tau_n}=\frac{e^{\mu_\rs}n^*_{\sX,\mu_\rs}}{p_{\mu_\rs}}-\frac{s_{\sX,0}}{\tau_0}-\frac{\mu_\rs s_{\sX,1}}{\tau_1},
\] we further get,
\begin{multline*}
\frac{ e^{\mu_\rd}  n^*_{\sX,\mu_\rd}}{p_{\mu_\rd}}-\frac{ e^{\mu_\rdd} n^*_{\sX,\mu_\rdd}}{p_{\mu_\rdd}} 
\leq \frac{(\mu_\rd-\mu_\rdd)s_{\sX,1}}{\tau_1} \\+\frac{\mu_\rd^2-\mu_\rdd^2}{\mu_\rs^2}   
\left( \frac{e^{\mu_\rs}n^*_{\sX,\mu_\rs}}{p_{\mu_\rs}}-\frac{s_{\sX,0}}{\tau_0}-\frac{\mu_\rs s_{\sX,1}}{\tau_1}\right). 
\end{multline*}
Finally, solving for $s_{\sX,1}$ gives
\begin{multline} \label{A:eqn:5}
s_{\sX,1} \geq \frac{\mu_\rs \tau_1 }{\mu_\rs(\mu_\rd-\mu_\rdd)-(\mu_\rd^2-\mu_\rdd^2)}
\Bigg[  \frac{ e^{\mu_\rd}  n^*_{\sX,\mu_\rd}}{p_{\mu_\rd}} \\ -\frac{ e^{\mu_\rdd} n^*_{\sX,\mu_\rdd}}{p_{\mu_\rdd}} + \frac{\mu_\rd^2-\mu_\rdd^2}{\mu_\rs^2} \left(    \frac{s_{\sX,0}}{\tau_0}-\frac{e^{\mu_\rs}n^*_{\sX,\mu_\rs}}{p_{\mu_\rs}} \right)                \Bigg]. 
\end{multline}

\subsubsection{Upper-bound on the number of single-photon errors }
An upper-bound on the number of single-photon errors can be obtained with just $m^*_{\sX,\mu_\rd}$ and  $m^*_{\sX,\mu_\rdd}$, i.e., by taking $e^{\mu_\rd}m^*_{\sX,\mu_\rd}/p_{\mu_\rd}-e^{\mu_\rdd}m^*_{\sX,\mu_\rdd}/p_{\mu_\rdd}$, it is easy to show that
\be \label{A:eqn:6}
v_{\sX,1} \leq\frac{\tau_1}{\mu_\rd-\mu_\rdd}\left( \frac{e^{\mu_\rd}  m^*_{\sX,\mu_\rd}}{p_{\mu_\rd}}       -\frac{e^{\mu_\rdd}  m^*_{\sX,\mu_\rdd}}{p_{\mu_\rdd}}   \right)      .
\ee

\subsubsection{Finite-size decoy-state analysis}
The bounds given above are still not applicable to the observed statistics since Eqs.~\eqref{A:eqn:4},~\eqref{A:eqn:5} and~\eqref{A:eqn:6} involve terms that are valid only in the asymptotic limit, i.e., $\{n_{\sX,k}^*\}_{k\in \cK}$ and $\{m_{\sX,k}^*\}_{k\in \cK}$. However, this is easily resolved by using Eqs.~\eqref{A:eqn:1} and~\eqref{A:eqn:2}. Specifically, let 
\begin{eqnarray} \label{A:eqn:7}
 n_{\sX,k}^* &\leq& n_{\sX,k}+\delta(n_\sX ,\eps_1)=:\tilde{n}_{\sX,k}^+, \\ \label{A:eqn:8}
 n_{\sX,k}^* &\geq& n_{\sX,k}-\delta(n_\sX ,\eps_1)=:\tilde{n}_{\sX,k}^- , 
\end{eqnarray} and
\begin{eqnarray} \label{A:eqn:9}
 m_{\sX,k}^* &\leq& m_{\sX,k}+\delta(m_\sX ,\eps_2)=:\tilde{m}_{\sX,k}^+, \\ \label{A:eqn:10}
 m_{\sX,k}^* &\geq& m_{\sX,k}-\delta(m_\sX ,\eps_2)=:\tilde{m}_{\sX,k}^- ,.
\end{eqnarray} for all values of $k$. Putting them into Eqs.~\eqref{A:eqn:4},~\eqref{A:eqn:5} and~\eqref{A:eqn:6}, we thus have the formulas as stated in the main text. 

\subsection{Secrecy analysis}
The secrecy analysis roughly follows along the lines of Ref.~\cite{Tomamichel2012a}, i.e., we use a certain family of entropic uncertainty relations to establish bounds on the smooth min-entropy of the raw key conditioned on Eve's information. 

To start with, let system $\rv{E}'$ be the information that Eve gathers on $\rv{X}_{\rm A}$, i.e., the raw key of Alice, up to the error-verification step. By applying privacy amplification with two-universal hashing~\cite{RennerThesis}, a $\esec$-secret key of length $\ell$ can be extracted from $\rv{X}_{\rm A}$. Specifically, the secret key is $\esec$-secret if $\ell$ is chosen such that
\be \label{A:eqn:11}
\ell = \left\lfloor \HminOp^{\nu}\left(\rvX_{\rm A}|\Eve'\right)-2\log_2\frac{1}{2\overline{\nu}} \right\rfloor,
\ee 
for $\nu+\overline{\nu} \leq \esec$ where $\nu,\overline{\nu}$ are chosen to be proportional to $\esec/(1-p_{\rm abort})$.~Here, $\HminOp^{\nu}\left(\rvX_{\rm A}|\Eve'\right)$ is the conditional smooth min-entropy, which quantifies the amount of uncertainty system $\rv{E}'$ has on $\rv{X}_{\rm A}$. In fact, this quantity is the heart of our security analysis. In the following, we show how to bound $\HminOp^{\nu}\left(\rvX_{\rm A}|\Eve'\right)$ using statistics obtained in the protocol. 

First, using a chain-rule inequality for smooth entropies, and the fact that $\leak$-bits and $\log_2 2/\epscorr$-bits of information were published during the error-correction and error-verification steps, respectively, we get $\HminOp^{\nu}\left(\rvX_{\rm A}|\Eve'\right)\geq \HminOp^{\nu}\left(\rvX_{\rm A}|\Eve\right)-\leak-\log_2 2/\epscorr$, where system $\rv{E}$ is the remaining (possibly quantum) information Eve has on $\rv{X}_{\rm A}$. In general, $\leak$ should be determined by the amount of leakage the actual protocol reveals during the error-correction step.  

Second, we decompose $\rv{X}_{\rm A}$ into $\rvX^\tn{v}_{\rm A}\rvX^{\tn{s}}_{\rm A}\rvX^{\tn{m}}_{\rm A}$, which are the corresponding bit-strings due to the vacuum, single-photon and multi-photon events. Note that this decomposition is known to Eve, i.e., the decomposition information is included inside system $\rv{E}$. By using a generalized chain-rule result from Ref.~\cite{Vitanov2013}, we have that
\begin{multline*}
\HminOp^{\nu}\left(\rvX_{\rm A}|\Eve\right) 
 \geq  \HminOp^{\alpha_1}\left(\rvX^{\tn{s}}_{\rm A}|\rvX^\tn{v}_{\rm A}\rvX^{\tn{m}}_{\rm A}\Eve\right)  
 \\+ \HminOp^{\alpha_3+2\alpha_4+\alpha_5}\left(\rvX^\tn{v}_{\rm A}\rvX^{\tn{m}}_{\rm A}|\Eve\right) - 2\log_2\frac{1}{\alpha_2}-1,
\end{multline*} for $\nu=2\alpha_1+\alpha_2+(\alpha_3+2\alpha_4+\alpha_5)$ where $ \alpha_i > 0$ for all $i$. Next, we use the same chain-rule again on the second term on the r.h.s. to get 
\begin{eqnarray*} 
\HminOp^{\alpha_3+2\alpha_4+\alpha_5}\left(\rvX^\tn{v}_{\rm A}\rvX^{\tn{m}}_{\rm A}|\Eve\right) &&
 \\ \geq   \HminOp^{\alpha_4}\left(\rvX^{\tn{m}}_{\rm A}|\rvX^\tn{v}_{\rm A}\Eve\right) & +&\HminOp^{\alpha_5}\left(\rvX^\tn{v}_{\rm A}|\Eve\right) 
 -2\log_2\frac{1}{\alpha_3}-1 
\\ \geq s_{\sX,0}-&2&\log_2\frac{1}{\alpha_3}-1. 
\end{eqnarray*} 
To get the second inequality, we used $\HminOp^{\alpha_4}\left(\rvX^{\tn{m}}_{\rm A}|\rvX^\tn{v}_{\rm A}\Eve\right)\geq 0$ and $\HminOp^{\alpha_5}\left(\rvX^\tn{v}_{\rm A}|\Eve\right) \geq \HminOp\left(\rvX^\tn{v}_{\rm A}|\Eve\right)=\HminOp \left(\rvX^\tn{v}_{\rm A}\right)=\log_2 2^{s_{\sX,0}}=s_{\sX,0} $. The former is given by the fact that all multi-photon events are taken to be insecure, i.e., due to the photon-number-splitting attack. The latter is based on the assumption that vacuum contributions contain zero information about the chosen bit values and the bits are uniformly distributed. 

Third, we provide a bound on the remaining smooth min-entropy quantity which is now restricted to the single-photon events, i.e., via the uncertainty relation for smooth entropies~\cite{Tomamichel2010S}. Under the assumption that Alice prepares the states using mutually unbiased bases (i.e., $\sX$ is the computational basis and $\sZ$ is the Hadamard basis), we can further bound this quantity with the max-entropy between Alice and Bob, which is directly given by the amount of correlation between them~\cite{Tomamichel2012a}. More precisely, we have
\begin{eqnarray*}
\HminOp^{\alpha_1}\left(\rvX^{\tn{s}}_{\rm A}|\rvX^\tn{v}_{\rm A}\rvX^{\tn{m}}_{\rm A}\Eve\right)  
&\geq& s_{\sX,1}-\HmaxOp^{\alpha_1}\left(\rvZ^{\tn{s}}_{\rm A}|\rvZ^\tn{s}_{\rm B}\right)
\\&\geq& s_{\sX,1}\left[1-h\left(\frac{c_{\sX,1}}{s_{\sX,1}}\right)\right],
\end{eqnarray*}
where the first inequality is given by the uncertainty relation~\cite{Tomamichel2010S} and the smooth max-entropy $\HmaxOp^{\alpha_1}\left(\rvZ^{\tn{s}}_{\rm A}|\rvZ^\tn{s}_{\rm B}\right)$ is a measure of correlations between $\rvZ^{\tn{s}}_{\rm A}$ and $\rvZ^{\tn{s}}_{\rm B}$. Here, $\rvZ^{\tn{s}}_{\rm A}$ and $\rvZ^{\tn{s}}_{\rm B}$ are the bit strings Alice and Bob would have obtained if they had measured in the basis $\sZ$ instead. The second inequality is achieved by using $\HmaxOp^{\alpha_1}\left(\rvZ^{\tn{s}}_{\rm A}|\rvZ^\tn{s}_{\rm B}\right)\leq s_{\sX,1}h(c_{\sX,1}/s_{\sX,1})$ (see \cite[Lemma~3]{Tomamichel2012a}), where $c_{\sX,1}$ is the number of phase errors in the single-photon events. Here, the number of phase errors $c_{\sX,1}$ has to be estimated via a random-sampling theory (without replacement) as these errors are not directly observed in the protocol. More concretely, by using a random-sampling without replacement result given in Ref.~\cite{Fung2010} which is based on an approximation technique for the hyper-geometric distribution, we have with probability at least $1-\alpha_1$,
\be
  \frac{c_{\sX,1}}{s_{\sX,1}} \leq \frac{v_{\sZ,1}}{s_{\sZ,1}} + \gamma\left(\alpha_1,  \frac{v_{\sZ,1}}{s_{\sZ,1}},s_{\sZ,1},s_{\sX,1} \right), \ee
where 
\[
\gamma\left(a,b,c,d \right):= \sqrt{\frac{(c+d)(1-b)b}{cd\log2}\log_2\left( \frac{c+d}{cd(1-b)b} \frac{1}{a^2}\right)}.
\]

Fourth, putting everything together, we arrive at a secret key length of
\begin{multline}
\ell=\bigg\lfloor s_{\sX,0}+s_{\sX,1}\left[1-h\left(\frac{c_{\sX,1}}{s_{\sX,1}}\right)\right]-\leak 
 -\log_2\frac{2}{\epscorr \beta} \bigg\rfloor,
\end{multline}
where~$\beta:=(\alpha_2\alpha_3\overline{\nu})^2$.~Note that~$s_{\sX,0}$, $s_{\sX,1}$, $s_{\sZ,0}$, $s_{\sZ,1}$, $v_{\sZ,1}$~are to be bounded by  Eqs.~\eqref{A:eqn:4}-\eqref{A:eqn:6} using the relations given by Eqs.~\eqref{A:eqn:7}-\eqref{A:eqn:10}.

Finally, after composing the error terms due to finite-sample sizes and setting $\alpha_4=\alpha_5=0$, the secrecy is
\be
\esec=2\left[2\alpha_1+\alpha_2+\alpha_3\right]+\overline{\nu}+10\eps_1+2\eps_2.
\ee
To get the secrecy given in the main text we set each error term to a common value $\eps$, thus $\esec=21\eps$.

\bibliographystyle{unsrt}

\end{document}